# Optically discriminating carrier-induced quasiparticle band gap and exciton energy renormalization in monolayer MoS$_2$

Kaiyuan Yao[1,2], Aiming Yan[3,4], Salman Kahn[3], Aslihan Suslu[5], Yufeng Liang[1], Edward S. Barnard[1], Sefaattin Tongay[5], Alex Zettl[3,4,6], Nicholas J. Borys[1,7*], and P. James Schuck[1,8*]

[1]Molecular Foundry Division, Lawrence Berkeley National Laboratory, Berkeley, CA 94720, USA
[2]Department of Mechanical Engineering, University of California, Berkeley, CA 94720, USA
[3]Department of Physics, University of California, Berkeley, CA 94720, USA
[4]Materials Sciences Division, Lawrence Berkeley National Laboratory, Berkeley, CA, 94720, USA
[5]Department of Materials Science and Engineering, Arizona State University, Tempe, AZ, 85287, USA
[6]Kavli Energy NanoSciences Institute at the University of California, Berkeley and the Lawrence Berkeley National Laboratory, Berkeley, CA, 94720, USA
[7]Department of Physics, Montana State University, Bozeman, MT, 59717, USA
[8]Department of Mechanical Engineering, Columbia University, New York, NY, 10027, USA

*njborys@lbl.gov; *pjschuck@lbl.gov

**Abstract**

Optoelectronic excitations in monolayer MoS$_2$ manifest from a hierarchy of electrically tunable, Coulombic free-carrier and excitonic many-body phenomena. Investigating the fundamental interactions underpinning these phenomena – critical to both many-body physics exploration and device applications – presents challenges, however, due to a complex balance of competing optoelectronic effects and interdependent properties. Here, optical detection of bound- and free-carrier photoexcitations is used to directly quantify carrier-induced changes of the quasiparticle band gap and exciton binding energies. The results explicitly disentangle the competing effects and highlight longstanding theoretical predictions of large carrier-induced band gap and exciton renormalization in 2D semiconductors.

**Introduction**

Monolayer transition metal dichalcogenide (ML-TMDC) semiconductors are exquisite optoelectronic materials that synergize the effects of strong confinement [1-3], intense many-body interactions [4-6] and spin-coupled valley degrees of freedom [7] in a robust, atomically-thin semiconductor with extended two-dimensional (2D) crystalline order. In ML-TMDCs, electronic excitations are collective phenomena that are described by a quasiparticle band structure which condenses the excitations into particles with momentum and energy that reflect the underlying many-body physics and crystal structure [8, 9]. The energetic separation between the quasiparticle valence and conduction bands, termed the 'quasiparticle band gap' or simply the 'band gap', governs the electronic properties in ML-TMDCs such as transport, formation of Ohmic contacts and band alignment in heterostructures [10-14]. Meanwhile, photoexcitations, which are essential to optoelectronic functionality [15-19], create electron-hole pairs within the quasiparticle band structure, forming a rich manifold of bound exciton states. The lowest-energy exciton – a strong dipole transition in these materials – determines the 'optical band gap' (i.e., the energetic threshold of optical absorption, sometimes termed the 'excitonic band gap'), which is energetically smaller than the quasiparticle band gap because of the electron-hole binding energy [20-22]. Strong physical and dielectric confinement make Coulombic interactions central to determining these quasiparticle and optical bandgaps, and an incredibly compelling aspect of ML-TMDCs is the ease by which the strength of this interaction can be manipulated, providing an unprecedented tunability of the quasiparticle and exciton energies [23-27].

In this letter, we experimentally disentangle and quantify the carrier-induced renormalization of both quasiparticle and optical band gaps in ML-MoS$_2$, providing a unified picture of these rich and complex effects in two-dimensional semiconductors. This quantification is enabled by the direct, all-optical identification of the carrier-density-dependent quasiparticle band gap using photoluminescence excitation (PLE) spectroscopy combined with steady state electrostatic gating to control the strength of the Coulombic interactions in the ML-MoS$_2$. Importantly, renormalization effects on the quasiparticle band gap and exciton binding energy tend to counteract each other, leading to only minimal changes in the optical band gap [28]. Thus, in conventional optical absorption spectroscopy, without direct identification of the quasiparticle band gap, quasiparticle and excitonic renormalization effects must be inferred from higher-lying excitonic states [25, 27, 29, 30]. Central to our approach, we demonstrate that the relative photoluminescence from defect-bound excitons (DXs) [31-33] diminishes with increased carrier doping and can identify the onset of photoexcitation of free carriers at the quasiparticle band gap. When combined with ground-state absorption and PLE spectroscopy, we can (1) track carrier-induced renormalization of the quasiparticle band gap and (2) fully deconvolve excitonic and quasiparticle renormalization effects. For both effects, we find renormalization of more than 150 meV over a moderate range of doping concentrations, agreeing remarkably well with previous theoretical predictions [23] and providing the first explicit experimental discrimination of the carrier-induced renormalization of the quasiparticle band gap from that of the exciton states in a quantum many-body system. Further, we observe that at low doping levels the band gap and exciton binding energy can be larger than 2.7 eV and 800 meV, respectively.

**Results and discussion**

Figure 1(a) shows a schematic of the PLE spectroscopy of back-gated ML-MoS$_2$ that reports the excitation-dependent photoluminescence as a function of free carrier density (MoS$_2$ grown on 285 nm SiO$_2$ on Si [34]). Complete experimental details are provided in the Supplemental Material [35], but we note that all measurements were performed at 80 K where radiative recombination from the DX states is activated [31]. At unbiased gating, ML-MoS$_2$ flakes were found to be heavily n-doped with a residual carrier concentration of $8.7 \times 10^{12}$ cm$^{-2}$, which likely results from interactions with the underlying substrate [29, 40]. Figure 1(b) shows the gate dependence of the relative intensities of the ground state excitonic emission at ~1.86 eV (A exciton) and of the DX states at ~1.72 eV. When unbiased (e.g., $V_g$=0 V), the PL of our samples is dominated by the trion [35]. Upon reducing the free carriers with increasingly negative gate voltages, the lower-energy emission from the DXs emerge: decreasing the concentration of free carriers increases the relative PL yield of the DX states with respect to the A excitonic emission. Previous work has shown that DX emission is due to excitons Coulombically bound to charged sulfur vacancy sites [31-33]. Here, we show that the relative balance of DX and A exciton emission depends upon carrier concentration, suggesting this Coulomb interaction is subject to carrier screening. A similar trend was also noted in ML-WSe$_2$ [25]. To quantify this effect, we estimate the relative yield of the DX emission by calculating the ratio of emission below 1.80 eV to the total emission. And as evidenced in Fig. 1b inset, which plots the ratio as a function of gate voltage, this experimental observable can be used to detect changes in the free carrier concentration in the system.

Figure 2 shows PLE spectroscopy of ML-MoS$_2$ at an intermediate carrier concentration, where emission intensities from the DXs and main A exciton states are comparable. In Fig. 2(a), the relative PL yield of the DX as a function of excitation energy is overlaid with the absorption spectrum, and the three prominent excitonic 'A', 'B' and 'C' absorption resonances are identified [1, 41]. Note first that the ratio of the DX emission to the A exciton emission generally increases with increasing excitation energy nearly in unison with the absorption from the higher-energy 'C' band [41, 42]. Secondly, a pronounced dip is observed at 2.07 eV, which nicely corresponds to the resonance energy of the 'B' exciton state. And finally, at 2.64 eV a small but pronounced decrease, deviating from the otherwise monotonic increase, is observed. Four individual PL spectra at representative energies are shown in Fig. 2(b) which exemplify the differences in the relative yield of DX emission.

The level diagram in Fig. 2(c) summarizes the absorption resonances and coupling pathways at these excitation energies in ML-MoS$_2$ [29, 41]. Although the C exciton is peaked at ~2.9 eV, its absorption resonance is broad, yielding a tail of closely spaced excited states that spans nearly to the optical band gap. The narrower resonant excitations of the A and B excitons are superimposed on the C exciton at ~1.9 and ~2.1 eV, respectively. At each excitation energy, a fraction of the C excitons can relax to form A excitons [29, 41] and DX states. The generally increasing trend of the DX emission yield vs. excitation energy indicates that the relative coupling of C excitons to DXs strengthens with increasing energy. Direct excitation of B excitons, on the other hand, enhances the *relative* number of A excitons, presumably because this additional set of absorbing states lies in the same region of the Brillouin zone and preferentially couples to the A exciton, decreasing the relative DX yield at 2.07 eV.

For the higher energy decrease in the DX yield at 2.635 eV, which is weaker than its lower energy counterpart at ~2.07 eV, neither a strong excitonic resonance exists [29, 30, 43, 44] nor are there any corresponding features in the absorption spectrum. Yet the PLE spectrum (Fig. 2(d)) of the total emission intensity exhibits a step-like increase at the same energy, which is well-described as the sum of a broad increasing background (from the C exciton tail) and a broadened step function (see Supplemental Material [35] for details), similar to a feature observed in our previous work [29]. Such a step-like increase in photoexcitation is anticipated for the absorption at the band edge of non-interacting electrons in two dimensions [8, 43, 44]. In conjunction with the decrease in the DX yield, we reason that this energy marks the onset of photoexcitation of the continuum of unbound electrons and holes [44, 45] near the quasiparticle band gap at the K/K' valleys. These unbound carriers reduce the emission yield of the DX states following the same mechanism as observed under electrostatic gating (Fig. 1(b)). The approximate reduction of defect PL yield at 2.64 eV is 2%. Using the linear trend fitted from Fig. 1(b), such a reduction corresponds to an injected carrier concentration of $3 \times 10^{11}$ cm$^{-2}$ which is on the same order as the estimated number of photoexcitations produced at these energies, of $\sim 8 \times 10^{10}$ cm$^{-2}$ (Supplemental Material [35]).

If our assertion is correct, the spectral signatures of direct excitation of the quasiparticle continuum in the PLE and relative DX yield should energetically shift with gate voltage as the quasiparticle band gap renormalizes [23]. Gate-dependent PLE and DX emission yield spectra are shown in Figure 3. At each gate voltage, the step feature in the PLE spectrum (Fig. 3(a)) and corresponding reduction in the relative emission yield of the DX (Fig. 3(b)) are observed (complete PLE data sets are shown in the Supplemental Material [35]). Using the PLE spectra alone, the energetic threshold for optical excitation of the continuum of unbound quasiparticles (i.e., the "continuum", $E_{con}$) is extracted from the position of a fitted step function (as described in Fig. 2) and are marked by the arrows in both the PLE and the relative DX emission yield spectra. Clearly, the two spectral features exhibit nearly identical renormalization effects. Starting at the residual doping concentration, $E_{con}$ first shifts to lower energies as the gate voltage decreases to -40 V and then reverses directions, shifting to higher energies as gate voltage further decreases to -90 V, which was our lowest obtainable gate voltage before dielectric breakdown. For positive gate voltages, the continuum excitation features rapidly diminish and are no longer clearly discernible possibly due to increased broadening of the continuum feature [45] and/or increased indirect optical absorption at higher carrier densities [46]. The strongly-correlated renormalization of the step feature in the PLE and the reduction in DX emission yield offer compelling evidence that these spectral features are indeed related to the quasiparticle band gap, and that their spectral shifts with gate voltage provide important insight into carrier-induced renormalization effects.

Notably, direct band-edge and excitonic transitions are expected to behave markedly differently in response to changes in carrier density [28, 45]. For example, the exciton absorption resonances renormalize by only ~10 meV [5], whereas the band gap energy is predicted to change by hundreds of meV [9] within a similar range of carrier concentration. To our knowledge, there are no excitonic states that are known or are predicted to renormalize over such a large energy range. In Figure 4, the renormalization of the quasiparticle band gap is quantified and compared to previous theoretical studies. The dependence of the continuum onset energy, $E_{con}$, on carrier concentration is summarized in Fig. 4(a) where the gate voltage has been converted to the electron concentration, $n_e$ (see Supplemental Material [35]). Careful distinction must now be drawn between the energetic onset of continuum excitations ($E_{con}$) and the quasiparticle band gap ($E_g$). In a doped system, $E_{con}$ is larger than $E_g$ due to Pauli blocking, as direct transitions can only occur from occupied states in the valence band to unoccupied states in the conduction band above the Fermi energy, $E_F$ (Fig. 4(a), inset). Using a parabolic approximation for the band extrema, $E_g$ is related to $E_{con}$ by $E_g = E_{con} - n_e \pi \hbar^2 / 2\mu q$, where $\hbar$ is the reduced Planck constant, $q$ is the electron charge, and $\mu$ is the exciton reduced mass [4]. From effective masses reported in literature [42], the quasiparticle band gap ($E_g$) at the residual doping level ($n_e = 8.7 \times 10^{12}$ cm$^{-2}$) is calculated to be $2.57 \pm 0.01$ eV where the uncertainty reflects the variations of multiple measurements. With decreasing electron concentration, the measured quasiparticle band gap increases nonlinearly, reaching $2.70 \pm 0.01$ eV at the lowest carrier concentration ($n_e = 1.8 \times 10^{12}$ cm$^{-2}$; $V_g = -90$ V) achieved in our measurements. By fitting a line to quasiparticle band gap $E_g$ at the four lowest electron concentrations, we estimate that $E_g$ of our samples at intrinsic doping concentrations is 2.78 ± 0.02 eV. Remarkably, the majority of the theoretical predictions of quasiparticle band gap from previous studies (Fig. 4(a); orange crosses) [23, 42, 47-49] are within 100 meV of our estimated value. We also note that the band gap we measured at the residual doping condition is comparable to recent photocurrent [50] and PLE [29, 30] measurements, but substantially higher than scanning tunneling microscopy (STM) measurements of ML-MoS$_2$ on conductive substrates [22, 51]. And further, a recent STM study of suspended ML-MoS$_2$ [52] finds a gap approaching the value we determined for the zero-doping condition.

Such a large, nonlinear renormalization of the quasiparticle band gap has previously been theoretically predicted and attributed to carrier-induced screening [23]. In Fig. 4(b), our experimental measurement of the quasiparticle band gap renormalization is compared to theoretical predictions [23] where $\Delta E_g$ denotes the change of the band gap from the residual doping concentration. We find that for the relative changes in the quasiparticle band gap, the experimental and theoretical results agree remarkably well. Moreover, the observed *band gap* renormalization of over 150 meV is more than one order of magnitude larger than any *excitonic* renormalization effects in ML-TMDCs [4, 25-27], further corroborating our assignment of the observed step feature in PLE spectra to be the continuum.

Finally, in Figure 5, the renormalization of the exciton binding energy is directly quantified by combining the PLE-derived values of the quasiparticle band gap and the optical band gap measured with gate-dependent absorption and PL spectra (Supplemental Material [35]). The extracted energies of the neutral A exciton ($A^0$) and charged A trion ($A^-$) states from absorption spectra are shown in Fig. 5(a). The corresponding carrier-dependent binding energy of $A^0$ can be calculated from its energetic separation from the quasiparticle band gap (Fig. 5(b)) and is found to be as large as $790 \pm 17$ meV at our lowest electron concentration. Extrapolating to lower concentrations, we estimate that the exciton binding energy at the zero-doping condition is $866 \pm 31$ meV, which is comparable to predictions by GW-BSE calculations [42, 48]. As the electron concentration is increased, the exciton binding energy rapidly decreases to $690 \pm 15$ meV at an electron concentration of $\sim 4.0 \times 10^{12}$ cm$^{-2}$ and then more gradually decreases to $660 \pm 12$ meV at the residual doping condition. This nonlinear behavior likely arises from the combined effects of increased Coulombic screening and phase space filling [53]. The resemblance between the renormalization trends for the quasiparticle band gap (Fig. 4(a)) and binding energy (Fig. 5(b)) reveals a linear relationship between these two (Fig. S9), similar to recent theoretical calculations predicting a general linear scaling law between exciton binding energy and quasiparticle band gap in 2D materials [54, 55]. We also note that carrier-induced effects on quasiparticle band gap and binding energy counteract each other, resulting in comparatively modest changes in excitonic transitions.

In conclusion, using the suppression of defect emission by free carriers in combination with PLE, PL and absorption spectroscopies, we have directly quantified carrier-induced quasiparticle and excitonic renormalization effects in gated ML-MoS$_2$ devices. At the lowest achieved doping level, the quasiparticle band gap is determined to be 2.70 ± 0.01 eV leading to an A exciton binding energy of 790 ± 17 meV. Both the quasiparticle band gap and binding energy renormalize by nonlinearly decreasing by over 150 meV as the electron concentration is increased to the residual doping level. Notably, our experimental results agree very well with previous theoretical predictions of the quasiparticle band gap [42, 47, 48] and renormalization effects [23]. As such, this spectroscopic approach serves as a facile way to identify the quasiparticle band gap in monolayer TMDC semiconductors in a broad range of device configurations, providing an all-optical compliment to STM [20, 22, 51, 52, 56]. For example, such information can be used in conjunction with ultrafast terahertz spectroscopy to study the rich many-body physics that govern exciton formation and coherence dynamics under both resonant and non-resonant excitation conditions [57-59]. Directly quantifying the fundamental quasiparticle band gap and exciton binding energies and their corresponding renormalization effects is essential for developing exciton-based optoelectronic devices in monolayer TMDC semiconductors that capitalize on their remarkable ability to tune the underlying many-body interactions.

**Acknowledgements**

The authors thank Scott Dhuey and Ed Wong for technical support. Work at the Molecular Foundry was supported by the Office of Science, Office of Basic Energy Sciences, of the U.S. Department of Energy under Contract No. DE-AC02-05CH11231. S. T. acknowledges funding from NSF DMR-1552220. This work was also supported by the Director, Office of Science, Office of Basic Energy Sciences, Materials Sciences and Engineering Division, of the U.S. Department of Energy under Contract No. DE-AC02-05-CH11231, which provided for crystal synthesis within the sp$^2$-Bonded Materials Program (KC2207) and preliminary sample characterization within the van der Waals Heterostructures Program (KCWF16); and by the National Science Foundation under Grant No. 1542741 which provided for device development.

# References

[1] K.F. Mak, C. Lee, J. Hone, J. Shan, and T.F. Heinz, Phys. Rev. Lett. **105**, 136805 (2010).

[2] A. Splendiani, L. Sun, Y. Zhang, T. Li, J. Kim, C.Y. Chim, G. Galli, and F. Wang, Nano Lett. **10**, 1271 (2010).

[3] W. Zhao, Z. Ghorannevis, L. Chu, M. Toh, C. Kloc, P.H. Tan, and G. Eda, ACS Nano **7**, 791 (2012).

[4] K.F. Mak, K. He, C. Lee, G.H. Lee, J. Hone, T.F. Heinz, and J. Shan, Nat. Mater. **12**, 207 (2013).

[5] E.J. Sie, A.J. Frenzel, Y.H. Lee, J. Kong, and N. Gedik, Phys. Rev. B, **92**, 125417 (2015).

[6] C.H. Lui, A.J. Frenzel, D.V. Pilon, Y.H. Lee, X. Ling, G.M. Akselrod, J. Kong and N. Gedik, Phys. Rev. Lett. **113**, 166801 (2014).

[7] K.F. Mak, K. He, J. Shan, and T.F. Heinz, Nat. Nanotechnol. **7**, 494 (2012).

[8] C.F. Klingshirn, *Semiconductor optics* (Springer Science & Business Media, 2012).

[9] L. Venema, B. Verberck, I. Georgescu, G. Prando, E. Couderc, S. Milana, M. Maragkou, L. Persechini, G. Pacchioni, and L. Fleet, Nat. Phys. **12**, 1085 (2016).

[10] S. Das, H.Y. Chen, A.V. Penumatcha, and J. Appenzeller, Nano Lett. **13**, 100 (2012).

[11] X. Cui, G.H. Lee, Y.D. Kim, G. Arefe, P.Y. Huang, C.H. Lee, D.A. Chenet, X. Zhang, L. Wang, F. Ye, F. Pizzocchero, B.S. Jessen, K. Watanabe, T. Taniguchi, D. A. Muller, T. Low, P. Kim, J. Hone, Nat. Nanotechnol. **10**, 534 (2015).

[12] C.H. Lee, G.H. Lee, A.M. Van Der Zande, W. Chen, Y. Li, M. Han, X. Cui, G. Arefe, C. Nuckolls, T.F. Heinz, J. Guo, J. Hone, and P. Kim, Nat. Nanotechnol. **9**, 676 (2014).

[13] R. Kappera, D. Voiry, S.E. Yalcin, B. Branch, G. Gupta, A.D. Mohite, and M. Chhowalla, Nat. Mater. **13**, 1128 (2014).

[14] Z. Lin, A. McCreary, N. Briggs, S. Subramanian, K. Zhang, Y. Sun, X. Li, N.J. Borys, H. Yuan, S.K. Fullerton-Shirey, A. Chernikov, H. Zhao, S. McDonnell, A. M. Lindenberg, K. Xiao, B.J. LeRoy, M. Drndić, et. al., 2D Mater. **3**, 042001 (2016).

[15] S. Wu, S. Buckley, J.R. Schaibley, L. Feng, J. Yan, D.G. Mandrus, F. Hatami, W. Yao, J. Vučković, A. Majumdar, and X. Xu, Nature (London) **520**, 69 (2015).

[16] Y. Ye, Z.J. Wong, X. Lu, X. Ni, H. Zhu, X. Chen, Y. Wang, and X. Zhang, Nat. Photonics **9**, 733 (2015).

[17] M. Bernardi, M. Palummo, and J.C. Grossman, Nano Lett. **13** 3664 (2013).

[18] C. Chakraborty, L. Kinnischtzke, K.M. Goodfellow, R. Beams, and A.N. Vamivakas, Nature nanotechnol. **10**, 507 (2015).

[19] F.H.L. Koppens, T. Mueller, P. Avouris, A.C. Ferrari, M.S. Vitiello, and M. Polini, Nature Nanotechnol. **9** 780 (2014).

[20] M.M. Ugeda, A.J. Bradley, S.F. Shi, H. Felipe, Y. Zhang, D.Y. Qiu, W. Ruan, S.K. Mo, Z. Hussain, Z.X. Shen, F. Wang, S. G. Louie, and M. F. Crommie, Nat. Mater. **13**, 1091 (2014).

[21] K. He, N. Kumar, L. Zhao, Z. Wang, K.F. Mak, H. Zhao, and J. Shan, Phys. Rev. Lett. **113**,

026803 (2014).

[22] C. Zhang, A. Johnson, C.L. Hsu, L.J. Li, and C.K. Shih, Nano Lett. **14**, 2443 (2014).

[23] Y. Liang, and L. Yang, Phys. Rev. Lett. **114**, 063001 (2015).

[24] A. Chernikov, C. Ruppert, H.M. Hill, A.F. Rigosi, and T.F. Heinz, Nat. Photonics **9**, 466 (2015).

[25] A. Chernikov, A.M. van der Zande, H.M. Hill, A.F. Rigosi, A. Velauthapillai, J. Hone, and T.F. Heinz, Phys. Rev. Lett. **115**, 126802 (2015).

[26] J.S. Ross, S. Wu, H. Yu, N.J. Ghimire, A.M. Jones, G. Aivazian, J. Yan, D.G. Mandrus, D. Xiao, W. Yao, and X. Xu, Nat. Commun. **4**, 1474 (2013).

[27] B. Liu, W. Zhao, Z. Ding, I. Verzhbitskiy, L. Li, J. Lu, J. Chen, G. Eda and K.P. Loh, Advan. Mater. **28**, 6457 (2016).

[28] S. Gao, Y. Liang, C.D. Spataru, and L. Yang, Nano Lett. **16**, 5568, (2016).

[29] N.J. Borys, E.S. Barnard, S. Gao, K. Yao, W. Bao, A. Buyanin, Y. Zhang, S. Tongay, C. Ko, J.Suh, A. Weber-Bargioni, J. Wu, Y. Li, P.J. Schuck, ACS Nano **11**, 2115 (2017).

[30] H.M. Hill, A.F. Rigosi, C. Roquelet, A. Chernikov, T.C. Berkelbach, D.R. Reichman, M.S. Hybertsen, L.E. Brus, and T.F. Heinz, Nano Lett. **15**, 2992 (2015).

[31] S. Tongay, J. Suh, C. Ataca, W. Fan, A. Luce, J.S. Kang, J. Liu, C. Ko, R. Raghunathanan, J. Zhou, F. Ogletree, J. Wu, et. al., Sci. Rep. **3**, 2657 (2013).

[32] P.K. Chow, R.B. Jacobs-Gedrim, J. Gao, T.M. Lu, B. Yu, H. Terrones, and N. Koratkar, ACS Nano **9**, 1520 (2015).

[33] V. Carozo, Y. Wang, K. Fujisawa, B. Carvalho, A. McCreary, S. Feng, Z. Lin, et al, Sci. Adv. **3**, 1602813 (2017).

[34] A. Yan, W. Chen, C. Ophus, J. Ciston, Y. Lin, K. Persson, and A. Zettl, Phys. Rev. B **93**, 041420 (2016).

[35] See https://journals.aps.org/prl/abstract/10.1103/PhysRevLett.119.087401 for Supplemental Material which includes Refs. [36-39].

[36] C. Robert, D. Lagarde, F. Cadiz, G. Wang, B. Lassagne, T. Amand, A. Balocchi, *et al*, Phys. Rev. B **93**, 205423 (2016).

[37] S.J. Orfanidis, *Electromagnetic waves and antennas* (New Brunswick, NJ, Rutgers University, 2002).

[38] M.L. Cohen, and S.G. Louie, *Fundamentals of Condensed Matter Physics* (Cambridge University Press, 2016).

[39] D. Sun, Y. Rao, G.A. Reider, G. Chen, Y. You, L. Brézin, *et al*, Nano Lett. **14**, 5625 (2014).

[40] B. Radisavljevic, A. Radenovic, J. Brivio, I.V. Giacometti, and A. Kis, Nat. Nanotechnol. **6**, 147 (2011).

[41] D. Kozawa, R. Kumar, A. Carvalho, K.K. Amara, W. Zhao, S. Wang, M. Toh, R.M. Ribeiro, A.C. Neto, K. Matsuda, and G. Eda, Nat. Commun. **5**, 4543 (2014).

[42] D.Y. Qiu, H. Felipe, S.G. Louie, Phys. Rev. Lett. **111**, 216805 (2013).

[43] A. Chernikov, T.C. Berkelbach, H.M. Hill, A. Rigosi, Y. Li, O.B. Aslan, D.R. Reichman, M.S. Hybertsen, and T.F. Heinz, Phys. Rev. Lett. **113**, 076802 (2014).

[44] G. Berghaeuser, E. Malic, Phys. Rev. B, **89**, 125309 (2014).

[45] A. Steinhoff, M. Rosner, F. Jahnke, T.O. Wehling and C. Gies, Nano Lett. **14**, 3743 (2014).

[46] A.E. Ruckenstein, and S. Schmitt-Rink, Phys. Rev. B **35**, 7551 (1987).

[47] T. Cheiwchanchamnangij, W.R. Lambrecht, Phys. Rev. B **85**, 205302 (2012).

[48] A. Ramasubramaniam, Phys. Rev. B **86**, 115409, (2012).

[49] H. Shi, H. Pan, Y.W. Zhang, and B.I. Yakobson, Phys. Rev. B **87**, 155304 (2013).

[50] A.R. Klots, A.K.M. Newaz, B. Wang, D. Prasai, H. Krzyzanowska, J. Lin, D. Caudel, N.J. Ghimire, J. Yan, B.L. Ivanov, and K.A. Velizhanin, Sci. Rep. **4**, 6608 (2014).

[51] J. Shi, M. Liu, J. Wen, X. Ren, X. Zhou, Q. Ji, D. Ma, Y. Zhang, C. Jin, H. Chen and S. Deng, Advan. Mater. **27**, 7086 (2015).

[52] N. Krane, C. Lotze, J.M. Läger, G. Reecht, and K.J. Franke, Nano Lett. **16**, 5163 (2016).

[53] C. Zhang, H. Wang, W. Chan, C. Manolatou, and F. Rana, Phys. Rev. B, **89**, 205436 (2014).

[54] J. Choi, P. Cui, H. Lan, and Z. Zhang, Phys. Rev. Lett. **115**, 066403 (2015).

[55] M. Zhang, L. Huang, X. Zhang, and G. Lu, Phys. Rev. Lett. **118**, 209701 (2017)

[56] X. Zhou, K. Kang, S. Xie, A. Dadgar, N.R. Monahan, X.Y. Zhu, J. Park, and A.N. Pasupathy, Nano Lett. **16**, 3148 (2016).

[57] M. Kira and S.W. Koch, Prog. Quant. Electro. **30**, 155 (2006).

[58] C. Bottge, S.W. Koch, L. Schneebeli, B. Breddermann, A. C. Klettke, M. Kira, B. Ewers, *et al*, Phys. Status Solidi (B) **250**, 1768 (2013).

[59] R. Ulbricht, E. Hendry, J. Shan, T.F. Heinz, and M. Bonn, Rev. Mod. Phys. **83**, 543 (2011).

# Figures

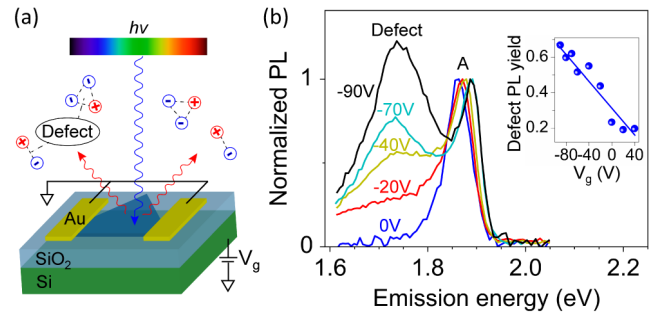

**Figure 1.** (a) Schematic of gate dependent photoluminescence excitation (PLE) spectroscopy on monolayer MoS$_2$. (b) Normalized PL spectra measured under different gate voltages with 2.5eV excitation. Inset shows the dependence of defect PL yield as a function of gate voltage.

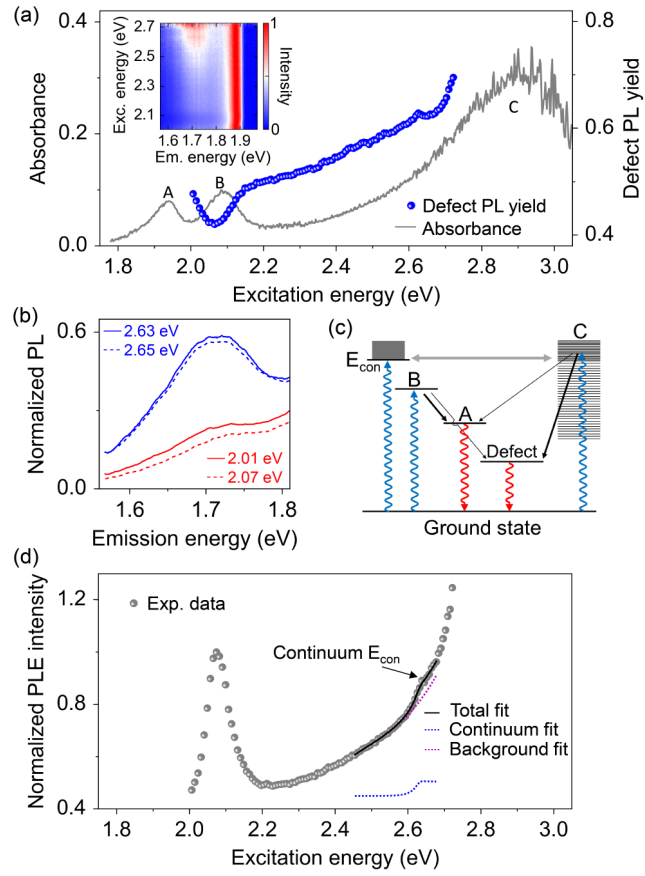

**Figure 2.** Identification of the quasiparticle band gap in monolayer MoS$_2$ from PLE spectroscopy. (a) Dependence of the yield of defect PL on excitation energy (blue dots), overlaid on the absorption spectrum (gray, taken from samples transferred to a quartz substrate). Inset shows the color contour of normalized PL spectra measured at different excitation energies. (b) Comparing the defect PL spectra (normalized to the A exciton; full spectra are shown in Supplemental Material [35]) under excitation energies that are on and off resonance of the continuum edge and B exciton. (c) Schematic level diagrams showing the relevant relaxation pathways of photo-generated excitations. A complete diagram is shown in Supplemental Material [35]. (d) Experimental PLE spectrum (gray dots) and total fit (black solid line) with contribution from the continuum (blue dotted line, with offset) and tail of the C exciton (magenta dotted line).

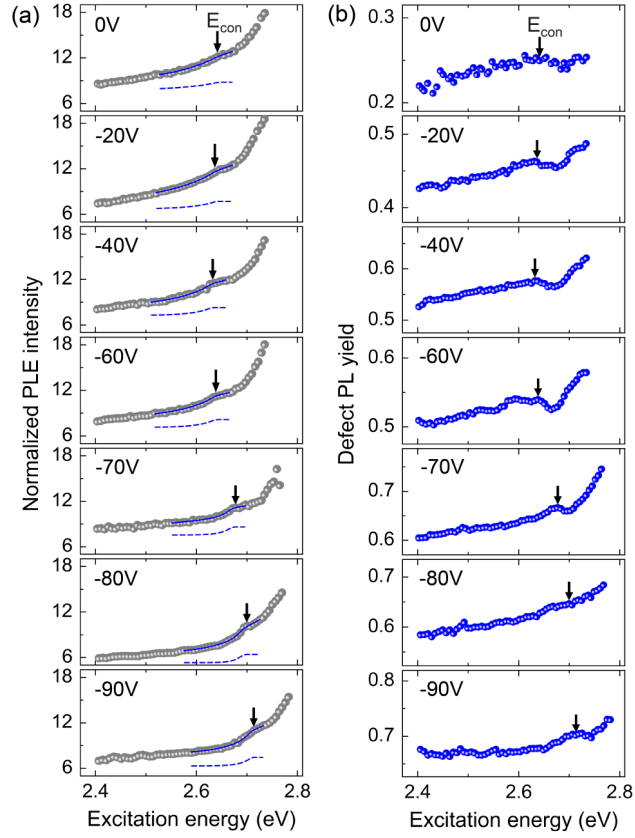

**Figure 3.** Gate dependent PLE spectroscopy of monolayer $MoS_2$. (a) PLE spectra of the integrated emission measured at different gate voltages. Experimental data, total fit, and the continuum contribution (with offset) are represented as gray dots, blue solid lines, and blue dashed lines, respectively. The PLE intensities are normalized to the oscillator strength (i.e., step height) of the fitted continuum function. (b) The excitation-energy dependent relative yield of defect PL at different gate voltages. The arrows in (a) and (b) represent the same energy of $E_{con}$ fitted from (a) as described in the text.

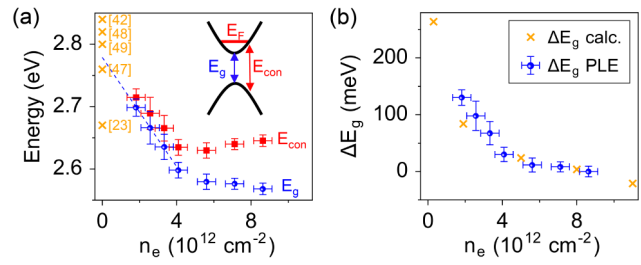

**Figure 4.** Carrier induced renormalization of the quasiparticle band gap of monolayer $MoS_2$. (a) Dependence on electron doping concentration $n_e$ of the measured continuum onset energy $E_{con}$ (red squares) and quasiparticle band gap $E_g$ (blue dots). Predicted quasiparticle band gap energies from previous studies ([23, 42, 47-49]) are also plotted for comparison (orange crosses). (b) Direct comparison of the measured *change* of quasiparticle band gap (blue dots) to previous theoretical predictions (orange crosses, [23]).

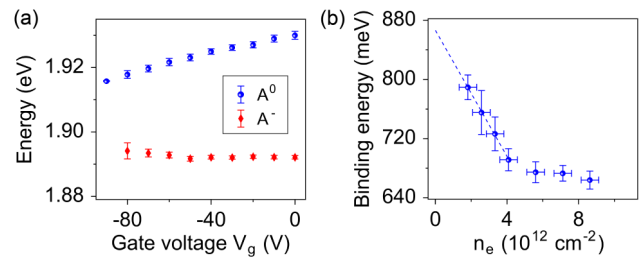

**Figure 5.** Carrier-induced renormalization of the exciton binding energy in monolayer MoS$_2$. (a) Dependence of the absorption energies of the neutral A exciton (blue dots) and charged A$^-$ trion (red diamonds) on gate voltage. (b) Renormalization of the binding energy of neutral A exciton with electron concentration.

# Supplemental Material

**Content**
- I. Material growth and device fabrication.
- II. Photoluminescence excitation spectroscopy and analysis.
- III. Identification of exciton and trion states from reflectivity spectra
- IV. Determination of gate induced doping concentrations from Stokes shift.

## I. Material growth and device fabrication

Monolayer MoS$_2$ (ML-MoS$_2$) was grown by chemical vapor deposition (CVD) [1]. The CVD growth was carried out in a two-zone tube furnace. Sulfur and MoO$_3$ precursors were placed inside a quartz tube and located in zone 1 and zone 2, respectively, of the tube furnace. The SiO$_2$/Si substrates for MoS$_2$ to grow on were placed on top of the crucible that held MoO$_3$ precursor in zone 2. N$_2$ gas was flown through the quartz tube during the entire growth process. The temperature for zone 1 and zone 2 was kept at 105 °C for 3 hours, and then ramped up to 517 °C (over 30 minutes) for zone 2 and stayed at 517 °C for 30 minutes while zone 1 was kept at 105 °C during this time. Before this point, the N$_2$ gas was flown at 200 sccm and was changed to 9 sccm afterwards. The temperatures for zone 1 and 2 were then increased to 400 °C and 820 °C, respectively, and stayed at these temperatures for 10 minutes. After this 10 minutes' growth was over, the tube furnace was turned off and cooled down naturally to room temperature.

The back-gated devices were fabricated by conventional optical lithography, followed by metal (Ti/Au 2nm/50nm) evaporation and lift-off. Electrical transport measurements reveal the typical n-type doping behavior at the residual condition, as shown in Fig. S1.

## II. Photoluminescence excitation spectroscopy and analysis

Photoluminescence excitation (PLE) spectroscopy was performed with a supercontinuum laser (5ps pulse width, 40MHz repetition rate) passed through an acoustic optical tunable filter as excitation source. The laser beam was expanded to approximate a parallel incident beam. Diameter of the illumination area was ~10μm. The laser power was kept very low (less than 5W/cm$^2$) to ensure that photoluminescence (PL) intensity scales linearly with excitation power. The PL spectra were collected with a cooled CCD.

Here we estimate the photo-generated excitation density as following. The typical laser power used for PLE is ~7μW measured at the back aperture of objective, corresponding to a photon pump fluence of $5.3 \times 10^{11}$cm$^{-2}$ for 2.65eV. The absorbance of ML-MoS$_2$ at this energy is estimated as 15% from Figure 2(a) of the main text. Then the density of photons absorbed by MoS$_2$ per pulse is estimated to be $8 \times 10^{10}$cm$^{-2}$. At our measurement temperature of 80K, carrier lifetime is expected to be much longer than the laser pulse width of 5ps, due to the rapid thermalization of photo-generated carriers, and the reduced efficiency of nonradiative processes [2]. Hence, the photoexcitation density at the quasiparticle band edge is estimated to be $8 \times 10^{10}$cm$^{-2}$ for our experiments.

For PLE measurements on back-gate devices, the 285nm-thick SiO$_2$ dielectric layer impose a strong interference effect in the range of our excitation energies. The reflected waves from the SiO$_2$/Si substrate carry different amplitudes and phases, interfering coherently with the incident wave. Therefore, the actual electromagnetic field intensity felt by the atomically thin MoS$_2$ on top of the SiO$_2$ is not proportional to the measured incident laser power. We account for this effect by calculating the interference correction factor which is defined as the ratio of the actual local electromagnetic field intensity felt by MoS$_2$ on the SiO$_2$/Si substrate over intensity of the incident field. The directly measured PLE spectra are then scaled per the interference correction factor so that the effective excitation photon flux is the same for each excitation energy. The calculation is performed by the transfer matrix method [3], with wavelength dependent indexes of Si and SiO$_2$. The optical path length of MoS$_2$ flake is much shorter than that of 285nm SiO$_2$, and we hence neglect its effect in interference correction. Fig. S2(a) shows the calculated interference correction factor. To verify this correction method, similar CVD-grown monolayer MoS$_2$ flakes were transferred from the growth substrate (SiO$_2$/Si) to thick quartz substrates for PLE measurements. The PLE spectrum measured on quartz is compared with our interference-corrected PLE spectrum measured on SiO$_2$/Si, as shown in Fig. S2(b). The larger noise is due to significantly decreased (estimated to be about two orders of magnitude) quantum yield after the transfer process. We see that the uncorrected PLE spectrum acquired on SiO$_2$/Si takes a declining trend due to the interference effect (red dots in Fig. S2(b)), which does not reflect the intrinsic material property, as detailed in our previous work [4]. After correction, the spectrum shows a similar growing trend as that acquired on the interference-free quartz substrate.

The experimental data of PLE intensity can be well fitted by an exponentially-tailed step-function plus a polynomial background. The absorption contribution from C exciton tail and the ensemble of high-energy Rydberg exciton states of the spin-split B series are accounted for by the polynomial background. The absorption due to the continuum of A band, $I_{cont.}(E_{ext})$, as a first approximation, may be expressed as:

$$I_{cont.}(E_{ext}) = \begin{cases} A_{cont.}, & E_{ext} > E_{con} \\ A_{cont.} e^{\frac{E_{ext}-E_{con}}{E_U}}, & E_{ext} < E_{con} \end{cases} \quad (1)$$

Here $E_{ext}$ is the excitation energy, $E_{con}$ is the continuum onset energy, $E_U$ describes the width of low energy tail, and $A_{cont.}$ is a proportionality factor representing oscillator strength of the continuum. The step-like function for $E_{ext} > E_{con}$ is the expected function for free carrier density of states of a non-interacting 2DEG [5, 6, 7, 8]. In a system with strong electron-hole interaction, the absorption lineshape due to free carriers near the quasiparticle band edge may deviate slightly from the simple step-function [8]. Precise modeling of the band edge absorption lineshape warrants future computational works. In this work, however, we find that a step-function with an exponential tail can adequately capture the experimental spectral features. The exponential tail possibly accounts for band edge disorder states and the ensemble of highly excited Rydberg exciton states of the A series close to the continuum edge. Equation (1) is further convolved with a Gaussian function with full width half magnitude (FWHM) of 10meV to account for possible heterogeneous broadenings.

The PLE spectra shown in Fig. 2 and Fig. 3 of the main text are constructed by integrating the total PL counts in the spectrometer collection range, including emission from both the free A excitons and the defect-bound DX excitons. In Fig. S3 we compare the PLE spectra obtained by integrating different portion of the PL spectra. Fig. S3(a) shows the color contour of absolute PL counts measured at each different excitation energies complimentary to the inset of Fig. 2(a) of the main text, where the PL counts have been normalized to the A emission peak in order to visualize the variation of relative defect PL yield with excitation energy. Fig. S3(b) compares the PLE spectra obtained by integrating the total PL counts (green dots), A exciton PL counts (magenta dots), and defect PL counts (yellow dots). All three spectra show identical features (B exciton resonance and the change of slope at ~2.63 eV) only with background differences.

### III. Identification of exciton and trion states from reflectivity spectra

Absorption spectra are obtained from reflectivity measurements. A Halogen lamp was used for illumination. A 100μm pinhole was used with a 60X objective (NA=0.6) to ensure that the collected signal comes from a localized area of ~2μm in diameter, much smaller than the typical flake size. The reflectance spectra from the flake $R_f$ and from the substrate (quartz or SiO$_2$/Si) $R_s$ are collected, and the reflectivity contrast was calculated as $R_{con} = \frac{R_f - R_s}{R_s}$. For samples on quartz substrate, the absorbance $A$ is directly proportional to $R_{con}$ as $R_{con} = \frac{4A}{n_s - 1}$ [9] where $n_s$ is index of substrate. For samples on SiO$_2$/Si substrate, Kramers-Kronig method [10] is employed for fitting the reflectivity ratio ($R_f/R_s$) spectra (Fig. S5(a)) and obtaining the absorption spectra (Fig. S5(b)). The absorption peak energies of exciton and trion states are determined by fitting gate-dependent absorbance with two Lorentzian functions representing the A$^0$ neutral exciton (blue dashed line) and negatively-charged A$^-$ trion (red dashed line), and a polynomial background representing the contribution from oscillators located at higher energies outside the fitting range [10]. A 10meV Gaussian broadening has been applied. The fitted resonance energies are shown in Fig. 5(a) of the main text. Note that the vertical error bars in Fig. 5(a) come from standard deviation of the fitting process.

### IV. Determination of gate induced doping concentration from Stokes shift

In this section, we discuss the determination of electron doping concentration at each gating voltage by measuring gate-dependent Stokes shift [10]. Stokes shift here is defined as the energy difference between the absorption peak and PL peak of the same exciton states, which is expected to be linearly proportional to the Fermi energy and carrier doping concentration in a 2D system [10].

The absorption peak energy for trion and excitons states have been determined as in Section III above. In PL measurements, the defect PL overlaps significantly with the direct A exciton PL peak, and cannot be well fitted with simple Gaussian profiles. Since the defect PL has a sublinear power dependence and saturates at higher excitation density [11], the excitation density is increased (~1000W/cm$^2$) to saturate defect PL for more accurate fitting of exciton energies. We note that this excitation density is still several orders below Mott transition [12], and the A exciton PL peak is not observed to shift or broaden between the relatively higher and lower excitation densities in our experiments. Fig. S6(a) shows the gate-dependent PL spectra and fitting results. Under all gate voltages, the major PL peak can be well fitted with one single Lorentzian function. Fig. S6(b) shows the fitted results of peak energies from PL and absorption measurements. The PL peak is red-shifted with increasing electron doping concentration, agreeing well with the previously reported trend of negatively charged trion A[10]. Fig. S6(c) shows the gate dependent Stokes shift, and compares the results for two different assignments (neutral exciton $A^0$ or negatively-charged trion $A^-$) of the PL peak. The neutral gate voltage (corresponding to zero doping) is determined as the linearly extrapolated point where the Stokes shift becomes zero. If the PL peak was assigned as from neutral exciton $A^0$, the neutral gate voltage would be -163V and the electron doping concentration at $V_g$=-90 V would be as high as ~6×10$^{12}$cm$^{-2}$. However, the trion feature in absorption spectra has already disappeared at $V_g$=-90 V (Fig. S5), indicating that the electron doping concentration at this gate voltage is already close to intrinsic [13]. Hence, based on the gate-dependent energy shift of the PL peak, as well as the gate-dependent absorption oscillator strength of $A^-$ trion versus $A^0$ exciton, we determine that the PL peak is from $A^-$ trion instead of neutral exciton $A^0$. PL from the neutral exciton $A^0$ is not appreciable, as a possible result of the high residual doping concentration, as well as the low measurement temperature which favors luminescence from the lower energy trion state [10]. With this assignment, the neutral gate voltage $V_{g0}$ is determined to be -114±6V. The electron concentration $n_e$ at each applied $V_g$ is then determined from the capacitance (per area) of the 285nm-thick SiO$_2$ dielectric layer $C_{ox}$ as $n_e$= $C_{ox}(V_g-V_{g0})$. The horizontal error bars in Fig. 4 and Fig. 5 of the main text come from the standard deviation of the fitted neutral gate voltage.

The binding energy of the ground state neutral exciton $A^0$ is determined as the energy difference between quasiparticle band gap $E_g$ and the energy $E_{A^0}$ (i.e., the optical band gap) required to create an $A^0$ exciton. We note that $E_{A^0}$ could ideally be obtained from the PL peak of $A^0$. However, as discussed above, the PL spectra we measured at low temperature is dominated by $A^-$ and the features of PL from $A^0$ is not appreciable. On the other hand, $A^0$ and $A^-$ can be well deconvolved from absorption spectra as in Fig. S5, and the measured absorption peak energy $E_{A^0}^{abs}$ of the $A^0$ neutral exciton state is expected to be only slightly higher than the optical band gap $E_{A^0}$ by the Stokes shift $E_{sts}$ [10]. Thus, the optical band gap is rigorously obtained as $E_{A^0} = E_{A^0}^{abs} - E_{sts}$. Hence the binding energy is calculated as $E_{bnd}$= $E_g$ - $E_{A^0}$ =$E_g$- $E_{A^0}^{abs}$ +$E_{sts}$, and shown in Fig. 5(b) of the main text as a function of electron concentration, where the error bars of $E_{bnd}$ sum up that of $E_g$, $E_{A^0}^{abs}$ and $E_{sts}$.

# References

[1] A. Yan, W. Chen, C. Ophus, J. Ciston, Y. Lin, K. Persson, and A. Zettl, Phys. Rev. B **93** (2016). Ref. [34] in main text.

[2] C. Robert, D. Lagarde, F. Cadiz, G. Wang, B. Lassagne, T. Amand, A. Balocchi, P. Renucci, S. Tongay, B. Urbaszek, and X. Marie, Phys. Rev. B **93**, 205423 (2016).

[3] S.J. Orfanidis, *Electromagnetic waves and antennas* (New Brunswick, NJ, Rutgers University, 2002.)

[4] N.J. Borys, E.S. Barnard, S. Gao, K. Yao, W. Bao, A. Buyanin, Y. Zhang, S. Tongay, C. Ko, J. Suh, A. Weber-Bargioni, J. Wu, Y. Li, P. J. Schuck, ACS Nano **11**, 2115 (2017). Ref. [29] in main text.

[5] C.F. Klingshirn, *Semiconductor optics* (Springer Science & Business Media, 2012). Ref. [8] in main text.

[6] A. Chernikov, T.C. Berkelbach, H.M. Hill, A. Rigosi, Y. Li, O.B. Aslan, D.R. Reichman, M.S. Hybertsen, and T.F. Heinz, Phys. Rev. Lett. **113**, 076802 (2014). Ref. [43] in main text.

[7] G. Berghaeuser, E. Malic, Phys. Rev. B **89**, 125309 (2014). Ref. [44] in main text.

[8] M.L. Cohen, and S.G. Louie, *Fundamentals of Condensed Matter Physics* (Cambridge University Press, 2016).

[9] D. Sun, Y. Rao, G.A. Reider, G. Chen, Y. You, L. Brézin, A.R. Harutyunyan, and T.F. Heinz, Nano Lett. **14**, 5625 (2014).

[10] K.F. Mak, K. He, C. Lee, G.H. Lee, J. Hone, T.F. Heinz, and J. Shan, Nat. Mater. **12**, 207, (2013). Ref. [4] in main text.

[11] S. Tongay, J. Suh, C. Ataca, W. Fan, A. Luce, J.S. Kang, J. Liu, C. Ko, R. Raghunathanan, J. Zhou, F. Ogletree, J. Wu, et. al., Sci. Rep. **3**, 2657 (2013). Ref. [31] in main text.

[12] A. Chernikov, C. Ruppert, H.M. Hill, A.F. Rigosi, and Heinz, T.F., Nat. Photonics **9**, 466 (2015). Ref. [24] in main text.

[13] A. Chernikov, A.M. van der Zande, H.M. Hill, A.F. Rigosi, A. Velauthapillai, J. Hone, and T.F. Heinz, Phys. Rev. Lett. **115**, 126802 (2015). Ref. [25] in main text.

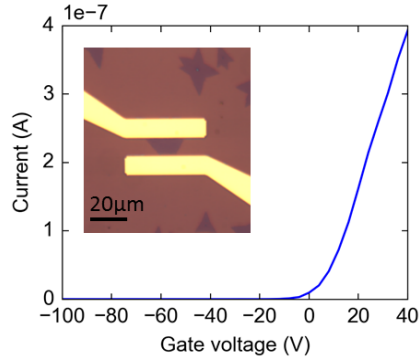

**Figure S1**. Typical transfer characteristics of fabricated monolayer MoS$_2$ field effect transistors measured at 80K. Inset shows the device photograph.

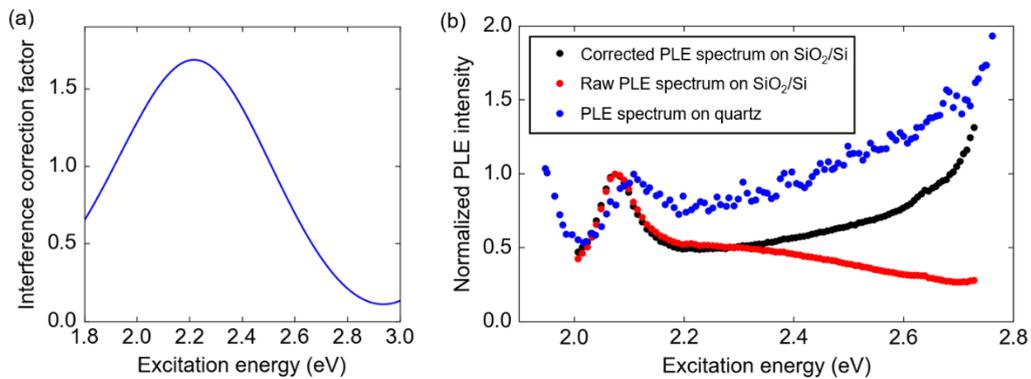

**Figure S2.** Correcting substrate induced thin film interference effect on PLE spectra of ML-MoS$_2$. (a) Calculated interference correction factor, as defined in the text. (b) Raw PLE spectrum measured for ML-MoS$_2$ on a 285nm-thick SiO$_2$/Si substrate (red dots), the interference-corrected PLE spectrum (black dots), and PLE spectrum measured for ML-MoS$_2$ transferred onto thick quartz substrate (blue dots). All three spectra are normalized to the peak of B exciton.

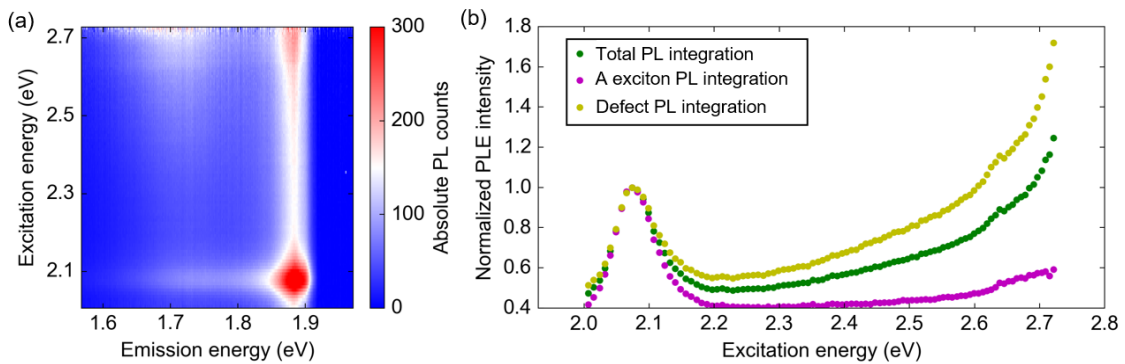

**Figure S3.** (a) Color contour of the absolute PL counts measured at each different excitation energies, with a $V_g$ of -60V. Note that this is the same data set as the inset of Fig.2b in the main text, but without normalization. (b) PLE spectra obtained by integrating the total PL counts (green dots), the A exciton PL counts (magenta dots),

and defect PL counts (yellow dots). All three spectra are normalized to their intensities at the B exciton peak at ~2.07eV.

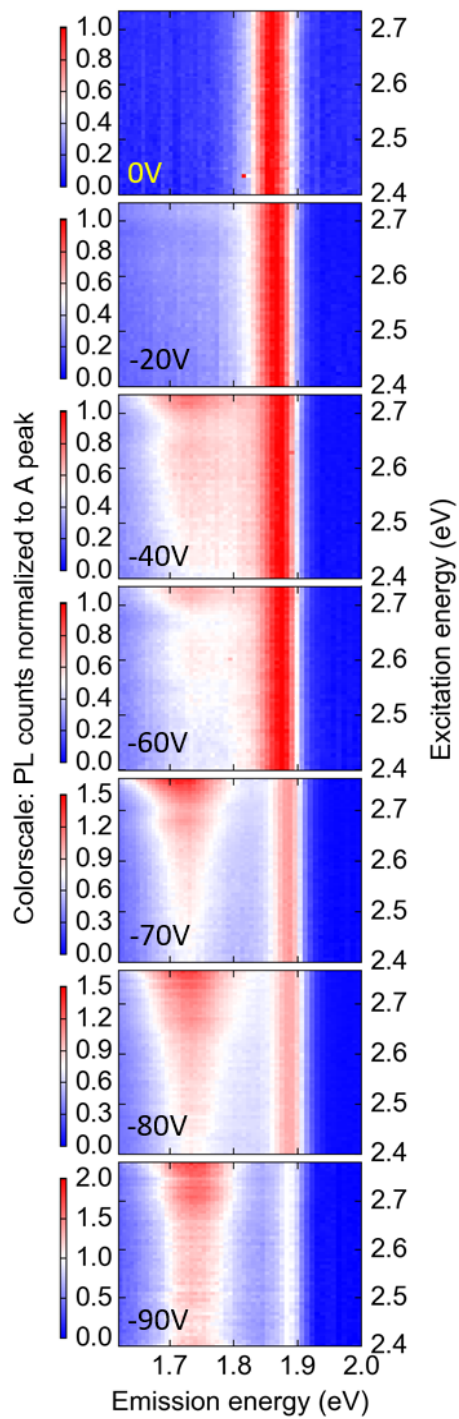

**Figure S4.** Color contour of normalized PL spectra measured with varying gate voltages, complementary to Fig. 3 of the main text.

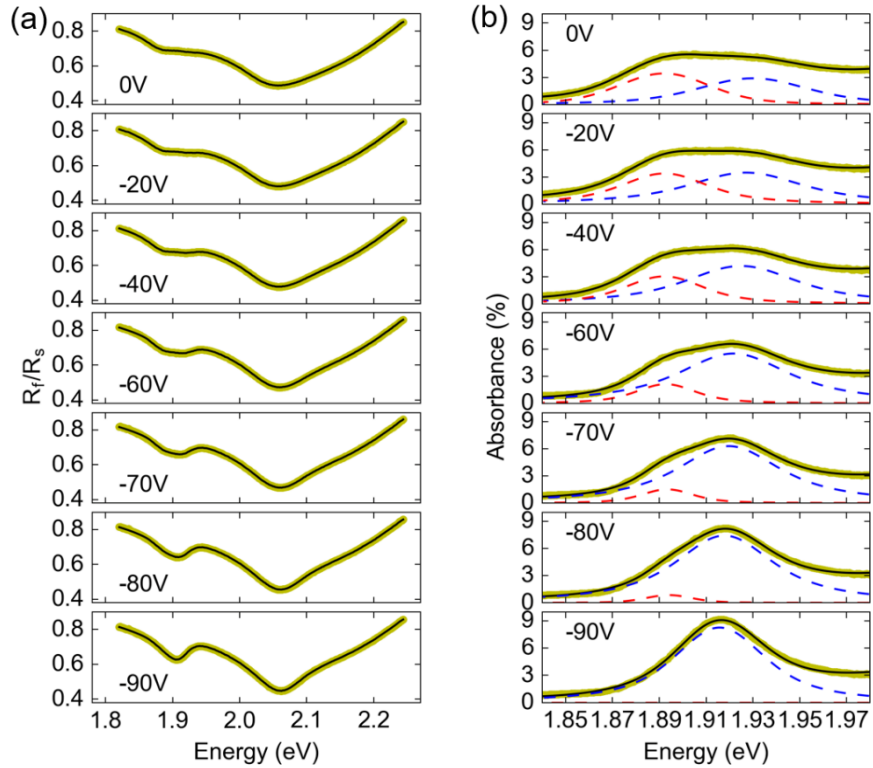

**Figure S5**. (a) The experimental data (yellow thick lines) of reflectance ratio $R_f/R_s$ is fitted by Kramers-Kronig method [10] (black thin lines) at different gate voltages as labeled. (b) The gate dependent absorption spectra (yellow thick lines) obtained from the Kramers-Kronig analysis. These absorption spectra are further fitted to identify exciton and trion states. Black thick lines show total fitting results, with blue (red) dashed lines representing the contribution from the $A^0$ neutral exciton (negatively-charged $A^-$ trion) states.

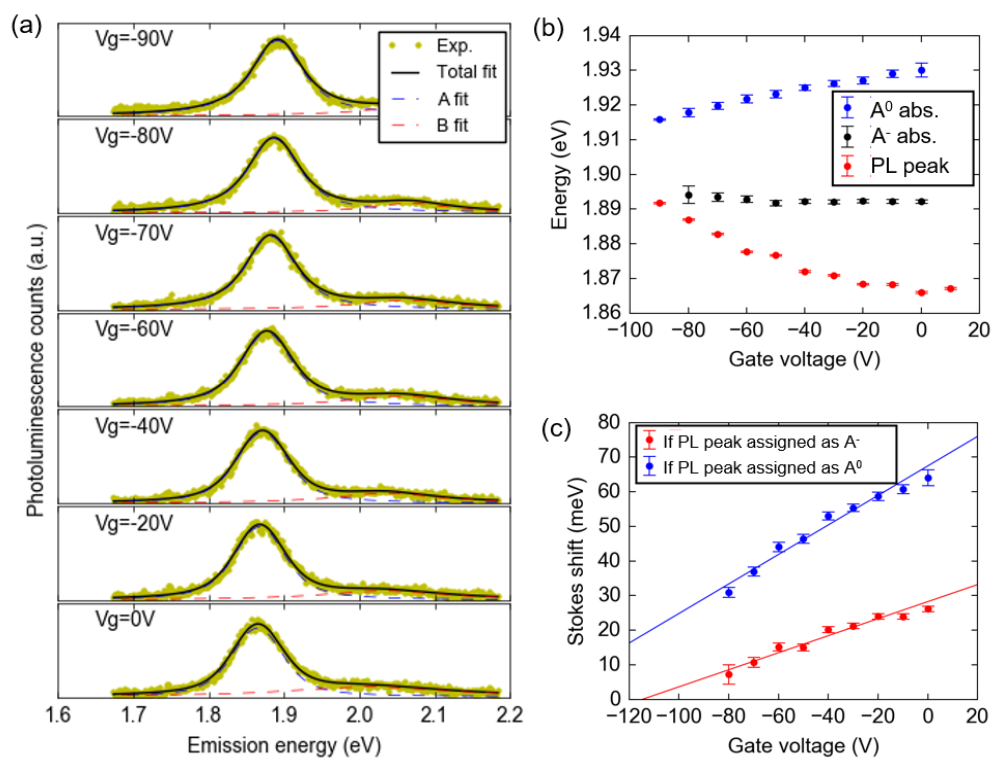

**Figure S6.** Determination of the neutral gate voltage that corresponds to zero electron doping concentration. (a) PL spectra measured at different gate voltages $V_g$, fitted with two Lorentzian functions representing the emission from A and B bands. (b) Dependence of the absorption peak energy of neutral exciton $A^0$ (blue dots), negatively-charged trion $A^-$ (black dots), and the PL peak energy (red dots) on gate voltage. (c) Dependence of Stokes shift on gate voltage, with two possible assignments of the PL peak. The assignment of $A^-$ state is favored (see text).

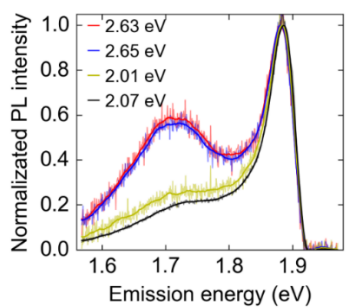

**Figure S7.** Full PL spectra showing the suppression of defect photoluminescence with resonant excitation of B exciton at ~2.07 eV and the continuum at ~2.65 eV, as complimentary to Figure 2 (b) of the main text. Semi-transparent thin lines represent the raw data. Solid thick lines show the smoothed spectra using a second order Savitzky Golay filter with a window size of 20 meV.

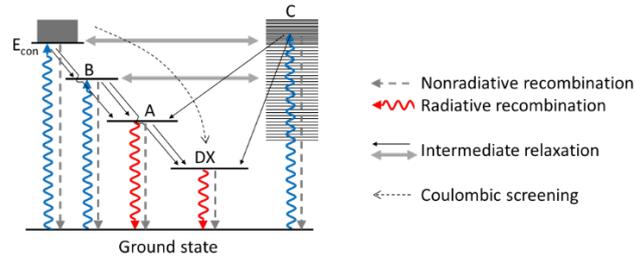

**Figure S8**. Complete level diagram as complimentary to Figure 2 (c) of the main text.

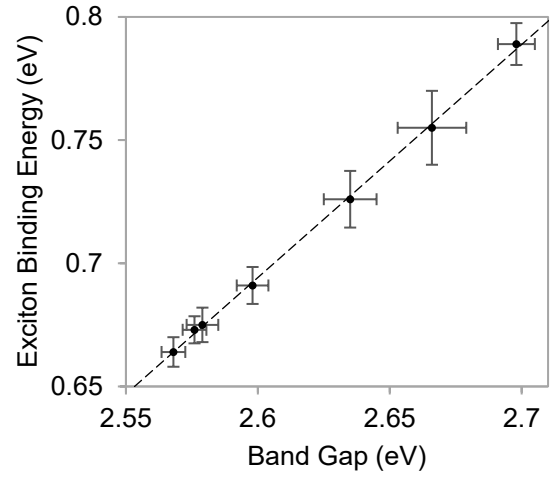

**Figure S9**. The exciton binding energies from Fig. 5b are plotted as a function of band gaps from Fig. 4a in the main text, showing a linear relationship as fitted by the dashed line.